# Acoustic particle detection – From early ideas to future benefits

Rolf Nahnhauer *

Deutsches Elektronen-Synchrotron, DESY, Platanenallee 6, D-15738 Zeuthen, Germany

A R T I C L E  I N F O

A B S T R A C T

The history of acoustic neutrino detection technology is shortly reviewed from the first ideas 50 years ago to the detailed R&D programs of the last decade. The physics potential of ultra-high energy neutrino interaction studies is discussed for some examples. Ideas about the necessary detector size and suitable design are presented.

© 2010 Elsevier B.V. All rights reserved.

## 1. The dawn of neutrino astroparticle physics

Very soon after the 1956 Science publication of Cowan et al. [1] about the discovery of neutrino interactions at the Savannah River reactor first ideas about experiments detecting neutrinos produced in the Earth atmosphere or in distant astrophysical sources appeared. In his talk "On high energy neutrino physics" at the 1960 Rochester conference [2] Markov stated: "in papers by Zhelesnykh[1] and myself (1958, 1960) possibilities of experiments with cosmic ray neutrinos are analyzed. We have considered those neutrinos produced in the Earth atmosphere from pion decay".

Reines [3] wrote in the same year in his article "Neutrino Interactions" for the Annual Review of Nuclear Science: "Interest in these possibilities{the study of cosmic neutrinos} stems from the weak interaction of neutrinos with matter, which means that they propagate essentially unchanged in direction and energy from their point of origin (…) and so carry information which may be unique in character".

The weak interaction of neutrinos makes them ideal messengers from processes inside massive galactic and extra-galactic objects up to the horizon of the universe, it has however the drawback that they are difficult to "catch" in detectors on Earth. An early idea how this could work was published by Greisen [4] in the same volume of Ann. Rev. Nucl. Science: "As a detector we propose a large Cherenkov counter about 15 m in diameter located in a mine far underground…. As fanciful though this proposal seems, we suspect that within the next decade cosmic ray neutrino detection will become one of the tools of both physics and astronomy".

It took more than a decade that indeed similar types of detectors were used to register the first extra-galactic neutrinos from the Supernova 1987A [5] at MeV energies.

## 2. First ideas about "particle sound"

Fifty years ago a well developed method to observe elementary particles in accelerator and cosmic ray experiments was to look for light they produced through Cherenkov radiation or scintillation effects when passing transparent media. At accelerators bubble chambers started their two decades lasting successful application, having their working principle still under study. In this context Askaryan [6] published in 1957 a paper about "Hydrodynamic radiation from the tracks of ionizing particles in stable liquids". He wrote "When ionizing particles pass through liquids … in addition, small micro-explosions due to localized heating, occur close to the tracks of the particles". This paper is mostly referenced as the first idea about acoustic particle detection. A very close application of this idea is used today in the Picasso experiment [7] searching for dark matter particles.

Early experimental tests of particle sound creation were done, e.g. 1969, by Beron and Hofstadter [8] using an electron beam at SLAC impinging on piezo-ceramic disks and by a group from the Kharkov university being able to show the linear dependence of the sound intensity on the deposited beam energy [9].

## 3. The blossom of acoustic particle detection

### 3.1. The DUMAND project

At the cosmic ray conference in Denver, 1973, a steering committee for the first deep under water detector for high energy neutrinos was formed. The idea for the DUMAND project (Deep Under water Muon and Neutrino Detector) was born and kept many people busy during the next two decades. A detailed description of its different phases can be found in Ref. [10]. From 1975 to 1980 many productive DUMAND workshops were organized in close collaboration of physicists from the US, Russia[2] and other countries.

---

* Tel.: +49 33762 77346; fax: +49 33762 77330.
E-mail address: rolf.nahnhauer@desy.de
[1] his student at that time.

[2] Roberts remembered 1992 [10]: "Russian participation in DUMAND was strong … and continued strong until it was abruptly cut off by the Reagan administration".

### 3.2. An acoustic detector for DUMAND and the Thermo-acoustic model

First thoughts about the addition of an acoustic detector to the optical telescope in the deep sea were discussed at the 1976 workshop in Hawaii [11,12]. There are also basic ideas for the "Thermo-acoustic Model" describing the creation of sound in neutrino interactions with nucleons were presented independently by Bowen [13] and Askaryan and Dolgoshein [11]. Demonstrative and detailed descriptions of the model can be found in Refs. [14,15].

The neutrino interacting with a nucleon of the target material produces an outgoing lepton and a hadronic particle cascade. In this process a large amount of energy is nearly instantaneously produced in a small volume of cascading particles. The overheating of that volume leads to a corresponding pressure pulse, which develops in a disk transverse to the incoming neutrino direction. The pressure amplitude is directly proportional to the cascade energy. In Ref. [14] a simple formula is given, which demonstrates the important quantities for the process:

$$p = (k/c_p)(E/R)M, \quad M = (f^2/2)(\sin x/x)$$
$$f = v_s/(2d), \quad x = (\pi L/2d)\sin \delta$$

with $p$ is the pressure amplitude, $E$ the cascade energy, $R$ the distance to receiver, $f$ the frequency, $v_s$ the speed of sound, $d$ the cascade diameter, $k$ the volume expansion coefficient, $c_p$ the specific heat, $L$ the cascade length, $\delta$ the angle between normal to cascade direction and receiver.

During the late 1970s several experiments took place to check the basic predictions of the Thermo-Acoustic Model [16,17]. Thermal expansion was identified as dominant mechanism for sound production; however, other contributions could not be excluded completely.

### 3.3. Thinking big

35 years ago it was not really clear how large an optical as well as an acoustic neutrino telescope would have to be in order to detect reasonable numbers of neutrinos. The main idea was: as big as possible, which led to impressive first design parameters as can be seen from Table 1. The numbers are close to those of today's largest neutrino telescopes under construction or in proposal phase.

The DUMAND project could not be finished successfully. By financial reasons the plan to build first a nine string detector was cut down to a three string project. In 1996 the project was finally terminated. Nevertheless it has prepared in many respects the ground on which later more successful experiments have been developed.

### 3.4. The acoustic desert and science fiction ideas

Attempts to build large acoustic detectors went, with some exceptions [18], for about 20 years to a dormant phase. However there were nevertheless optimists proposing to use directional neutrino beams from a sea-based accelerator, the "Geotron" to search for oil, gas or ores using movable acoustic detectors in the GENIUS (Geological Exploration by Neutrino Induced Underground Sound) project [19].

## 4. The revival of the acoustic particle detection technology

More than ten years ago controversial results from the study of the highest energy charged cosmic rays lead to the question of the existence of a cut-off in their energy spectra due to their interaction with the cosmic microwave background radiation [20]. A corresponding proof could be the detection of neutrinos from these interactions at energies above $10^{17}$ eV [21]. The observation and study of such neutrinos would give interesting information about several topics in particle and astrophysics [22,23] (see also Section 5). To detect the expected tiny neutrino flux, new techniques have to be developed to monitor the huge target volumes needed. Radio and acoustic detection methods came strongly in the focus of interest at that time again. New R&D programs at different sites and first radio experiments published results about neutrino flux limits. The status of the field was discussed at a series of workshops since then (see Table 2) with this one being the last in the chain.

In more than 20 different acoustic projects tasks were considered like sensor and detector design, calibration, signal processing, signal simulation, target material properties and environmental effects, in-situ test measurements, etc. Detailed information about all these studies is collected in the workshop proceedings [24–29]. In the following only a few examples can be mentioned in more detail.

### 4.1. Parasitic use of military acoustic arrays

Russian physicists used the AGAM antenna near Kamtchatka for acoustic particle detection studies [30]. It consists of 2400 hydrophones mostly sensitive below 2 kHz and should hear neutrino interactions above $10^{20}$ eV in a volume of 100 km$^3$. A project to use the portable submarine antenna MG-10M as a basic module of a deep water telescope has not been realized until now.

Using the French navy tracking array TREMAIL in the Mediterranean, ambient acoustic noise studies were done using eight hydrophones with 250 kHz sampling frequency in the early ANTARES project [31].

The first neutrino flux limit was given by the SAUND-I experiment using 7 hydrophones of the US AUTEC military array near the Bahamas [32]. An improved limit was published a few weeks after this workshop using a much larger array within the SAUND-II project [33].

In the UK the RONA array located in the North-West of Scotland provides eight hydrophones for a lot of studies of the ACORNE group about sensor calibration, signal filtering, noise reduction and source localization [34]. New results of the group are presented at this workshop [35].

### 4.2. Dedicated R&D arrays

At Lake Baikal a tetrahedral acoustic antenna with 4 hydrophones was placed in 150 m depth [36]. Mainly noise studies were done with this device, establishing low noise levels of a few mPa in the upper part of the lake. An update of this project is given at this workshop [37].

**Table 1**
Design parameters for early optical and acoustic neutrino telescopes [10,12].

|  | **Optical telescope** | **Acoustic detector** |
|---|---|---|
| **Volume** | 1.26 km$^3$ | 100 km$^3$ |
| **# strings** | 1261 | 10,000 |
| **x–y spacing** | 40 m | 100 m |
| **Depth** | 3900–4400 m | 3400–4400 m |
| **# sensors/string** | 18 | 10 |
| **Total # sensors** | 22,698 | 100,000 |

**Table 2**
Information about recent workshops about acoustic and radio neutrino detection.

| Time | Name | Location | Countries | Participants | Ref. |
|---|---|---|---|---|---|
| 2000 | RADHEP | Los Angeles | 6 | 50 | [24] |
| 2003 | Acou. mini-ws. | Stanford | 5 | 20 | [25] |
| 2005 | ARENA2005 | Zeuthen | 10 | 90 | [26] |
| 2006 | ARENA2006 | Newcastle | 9 | 50 | [27] |
| 2008 | ARENA2008 | Rome | 12 | 80 | [28] |
| 2010 | ARENA2010 | Nantes | 18 | 80 | [29] |

Long term noise studies were done with a similar device in the ONDE project in the deep Mediterranean sea near Sicily. The average noise in the {20:43} kHz band was found to be $5.4 \pm 2.2 \pm 0.3$ mPa [38]. The data provided also information about the number and habits of sperm whales at the site.

A second activity in the Mediterranean is the AMADEUS project within the ANTARES experiment [39]. 36 sensors at 6 storeys are placed at distances between 1 and 350 m. Noise conditions are similar to the ONDE results but strongly correlated with weather conditions. A signal location reconstruction leads to an angular distribution of marine sound sources. New results are presented at this conference [40].

The South Pole Acoustic Test Setup – SPATS – was built to evaluate the acoustic properties of the ice at the South Pole. Results are available already for the speed of sound of pressure and shear waves [41] and the sound attenuation length [42] which was measured to be about 300 m, i.e. more than an order of magnitude smaller than expected from theoretical estimates. A status report about recent achievements and future plans of the SPATS group [43] as well as results from a study of transient acoustic signals displaying a neutrino flux limit estimate [44] are given at this workshop.

## 5. Prospects of the acoustic technology

At the beginning of the last section it was already mentioned, that the observation of ultra-high energy neutrinos above $10^{17}$ eV would deliver interesting information for astrophysics, particle physics and cosmology. At these high energies only neutrinos allow to observe the whole universe whereas charged cosmic rays and photons are absorbed after less than ∼100 Mpc.

### 5.1. Neutrino sources

Three different types of sources are normally quoted to deliver neutrinos at energies above $10^{17}$ eV:

- Some models suggest that neutrinos produced in active galactic nuclei may have spectra with maxima at $10^{18}$ eV in a $E^2$ weighted flux distribution, however with lower intensities than originally expected [45].
- The most discussed neutrino source in the considered energy range comes from pion decay in the interaction of charged cosmic rays with the cosmic microwave background radiation (cosmogenic neutrinos) [20,21]. It's total flux is however uncertain by about three orders of magnitude and depends strongly on the chemical composition of the highest energy cosmic rays. If these are heavy ions like Fe, the flux will hardly be detectable [46]. Most experimental proposals use the more optimistic results from Ref. [47]. Present observations of high energy photon fluxes allow still observable number of neutrinos with reasonable detector sizes [48].
- There could exist even higher energetic neutrinos from the decay of topological defects [49]. These are super-heavy relic particles from the big bang with masses of $10^{21}$–$10^{25}$ eV. Neutrinos from their decay would carry about 5% of their energy.

### 5.2. Particle physics with UHE neutrinos

The neutrino nucleon cross-section is measured at accelerators up to ∼350 GeV. For the estimation of observable event numbers at ultra-high energies it has to be calculated at ten orders of magnitude higher energies. This is done using the Standard Model framework [50]. There are however several scenarios predicting large deviations from this extrapolation. In certain models the neutrinos may become strongly interacting above a threshold of ∼$10^{19}$ eV, with cross-sections increasing by more than five orders of magnitude (sphalerons [51], p-branes [52], string excitations [53]). Also models with black hole creation predict cross-sections enhanced by a factor 100 dependent on the number of extra-dimensions used [54].

If neutrinos with energies above $10^{21}$ eV exist, two long-standing problems of cosmology and particle physics could be tackled: the detection of the cosmic relic neutrino background radiation and the determination of neutrino masses [55,56]. By resonant annihilation of UHE neutrinos via the Z-resonance absorption, dips would appear in the measured neutrino spectra at $E_\nu^Z \approx (4 \times 10^{21} \text{ eV}/m_\nu)$ eV. As discussed in Ref. [57] one could even do neutrino mass spectroscopy with this method, being able at least to determine the correct mass hierarchy.

### 5.3. Test of basic symmetries

The violation of Lorentz invariance at the Planck scale: $E^2 = p^2 + m^2 + \eta(p^4/M_P^2)$ offers the possibility for neutrino splitting $\nu_A(p) \to \nu_A(p')\nu_B(q)\bar{\nu}_B(q')$ [58]. This effect leads e.g. to a cut-off of the cosmogenic neutrino spectrum depending on the $\eta$—parameter:

$$\eta < (E_{obs}/6 \times 10^{18} \text{ eV})^{-13/4}.$$

## 6. UHE-neutrino detector design today

All of the above discussed phenomena need at least more than a handful observed neutrinos to be studied with conclusive results. Taking into account the Standard Model cross-sections [50] and the modified flux [47], effective detector volumes have to be of the order of ∼1000 km$^3$ to fulfil the above requirement.

For such large target volumes it is hard to control the many possible background sources under control and to separated the few signals, if one uses only a single detection technology. A solution to this problem would be the multiple detection of the same neutrino signal with detection techniques with different systematic problems like the optical and/or radio and acoustic methods. This is possible in several target materials, presently probably best in ice [59]. In addition to a better background rejection one would get an improved energy and direction reconstruction and could cross-calibrate the different detector components.

## Acknowledgments

I would like to thank the organizers of the workshop for the invitation to give this talk.

I grateful acknowledge interesting discussions with C. Spiering about the subject who pointed me in particular to several historical details of neutrino astroparticle physics.